\documentclass[
superscriptaddress,
nofootinbib,
noeprint,
amsmath,amssymb,
aps,
prb,
twocolumn,
floatfix,
]{revtex4-1}

\usepackage{graphicx}
\usepackage{dcolumn}
\usepackage{bm}

\usepackage{epstopdf}
\usepackage{braket}
\usepackage[section]{placeins}

\begin{document}

\title{Quantum amplification of boson-mediated interactions}

\author{S. C. Burd}
\email{shaun.burd@colorado.edu}
\altaffiliation[Current address: ]{Department of Physics, Stanford University, Palo Alto, California 94305 USA}
\affiliation{Time and Frequency Division, National Institute of Standards and Technology, 325 Broadway, Boulder, Colorado 80305, USA}
\affiliation{Department of Physics, University of Colorado, Boulder, Colorado 80309, USA}

\author{R. Srinivas}
\affiliation{Time and Frequency Division, National Institute of Standards and Technology, 325 Broadway, Boulder, Colorado 80305, USA}
\affiliation{Department of Physics, University of Colorado, Boulder, Colorado 80309, USA}

\author{H. M. Knaack}
\affiliation{Time and Frequency Division, National Institute of Standards and Technology, 325 Broadway, Boulder, Colorado 80305, USA}
\affiliation{Department of Physics, University of Colorado, Boulder, Colorado 80309, USA}

\author{W. Ge}
\altaffiliation[Current address: ]{Department of Physics, Southern Illinois University, Carbondale, Illinois 62901 USA}
\affiliation{Texas A\&M University, College Station, Texas 77843, USA}

\author{A. C. Wilson}
\affiliation{Time and Frequency Division, National Institute of Standards and Technology, 325 Broadway, Boulder, Colorado 80305, USA}

\author{\break D. J. Wineland}
\affiliation{Time and Frequency Division, National Institute of Standards and Technology, 325 Broadway, Boulder, Colorado 80305, USA}
\affiliation{Department of Physics, University of Colorado, Boulder, Colorado 80309, USA}
\affiliation{Department of Physics, University of Oregon, Eugene, Oregon 97403, USA}

\author{D. Leibfried}
\affiliation{Time and Frequency Division, National Institute of Standards and Technology, 325 Broadway, Boulder, Colorado 80305, USA}

\author{J. J. Bollinger}
\affiliation{Time and Frequency Division, National Institute of Standards and Technology, 325 Broadway, Boulder, Colorado 80305, USA}

\author{D. T. C. Allcock}
\affiliation{Time and Frequency Division, National Institute of Standards and Technology, 325 Broadway, Boulder, Colorado 80305, USA}
\affiliation{Department of Physics, University of Colorado, Boulder, Colorado 80309, USA}
\affiliation{Department of Physics, University of Oregon, Eugene, Oregon 97403, USA}

\author{D. H. Slichter}
\email{daniel.slichter@nist.gov}
\affiliation{Time and Frequency Division, National Institute of Standards and Technology, 325 Broadway, Boulder, Colorado 80305, USA}

\date{\today}  

\maketitle

\textbf{Strong and precisely-controlled interactions between quantum objects are essential for quantum information processing~\cite{Bruzewicz2019,Blais2020}, simulation~\cite{Georgescu2014}, and sensing~\cite{Degen2017,Pezze2018}, and for the formation of exotic quantum matter~\cite{Bloch2008}.  A well-established paradigm for coupling otherwise weakly-interacting quantum objects is to use auxiliary bosonic quantum excitations to mediate the interactions. Important examples include photon-mediated interactions between atoms~\cite{Gerry2005}, superconducting qubits~\cite{Blais2004}, and color centers in diamond~\cite{Evans2018}, and phonon-mediated interactions between trapped ions~\cite{Cirac1995,Sorensen1999,Milburn2000} and between optical and microwave photons~\cite{Higginbotham2018}. Boson-mediated interactions can in principle be amplified through parametric driving of the boson channel; the drive need not couple directly to the interacting quantum objects.  This technique has been proposed for a variety of quantum platforms~\cite{Lu2015,Lemonde2016,Zeytino2017,Qin2018,Chen2019,Leroux2018,Arenz2018,Ge2019, Ge2019b, Groszkowski2020, Li2020}, but has not to date been realized in the laboratory.  Here we experimentally demonstrate the amplification of a boson-mediated interaction between two trapped-ion qubits by parametric modulation of the trapping potential~\cite{Ge2019}. The amplification provides up to a 3.25-fold increase in the interaction strength, validated by measuring the speedup of two-qubit entangling gates. This amplification technique can be used in any quantum platform where parametric modulation of the boson channel is possible, enabling exploration of new parameter regimes and enhanced quantum information processing.}

In many experimental platforms for quantum science, interactions between quantum objects are generated by coupling them via a shared auxiliary harmonic oscillator degree of freedom. The excitations of the harmonic oscillator are bosons (typically photons or phonons), which mediate interactions between the quantum objects. Such interactions have been used to demonstrate high-fidelity quantum logic gates~\cite{Ballance2016,Gaebler2016,McKay2019}, spin-squeezed states of atoms and ions~\cite{Meyer2001,Cox2016,Hosten2016a,Bohnet2016}, and the formation of novel phases of matter~\cite{mottl2012roton, leonard2017}.
 
 Achieving high fidelity generally requires that the effective interaction strength must dominate the characteristic rates of decoherence in the system.  Recent theoretical proposals\cite{Lu2015,Lemonde2016,Zeytino2017,Qin2018,Chen2019,Leroux2018,Arenz2018,Ge2019, Ge2019b, Groszkowski2020, Li2020} offer a way to increase the boson-mediated interaction strength through parametric modulation. When decoherence of the quantum objects to be coupled\textemdash including decoherence due to control fields used to implement the coupling~\cite{Ozeri2007}\textemdash is the primary source of infidelity, this technique can reduce that infidelity by decreasing the required interaction duration.   Stronger interactions could also increase speed and reduce control signal power requirements in large-scale quantum processors.

The physics of amplified boson-mediated interactions can be modeled by considering a set of quantum objects with associated operators $\hat{s}_{i}$ and a collective degree of freedom $\hat{S}\equiv\sum_{i}\beta_i\hat{s}_{i}$, with suitable coefficients $\beta_i$. This collective degree of freedom is coupled to a harmonic oscillator mode with annihilation and creation operators $\hat{a}$ and $\hat{a}^{\dagger}$ and frequency $\omega$, which can be parametrically modulated with characteristic strength $g$ at $2\omega+2\delta$ ($\delta$ is a system-dependent frequency offset). In a suitable interaction picture, the corresponding Hamiltonian takes the form~\cite{Ge2019} (see Methods) 

\begin{eqnarray}
\hat{H}_{M}&=&\frac{\hbar\Omega_{0}}{2}(\hat{S}^{\dagger}\hat{a}+\hat{S}\hat{a}^{\dagger})-
\hbar\delta\hat{a}^{\dagger}\hat{a} \nonumber\\
& &+\frac{\hbar g}{2}(\hat{a}^{2}e^{i\theta}+\hat{a}^{\dagger2}e^{-i\theta}),
\label{eq:H_M}
\end{eqnarray}
\noindent
where the first two terms describe the unamplified boson-mediated interactions, with coupling strength $\Omega_{0}$ and detuning $\delta$, and the third term describes the parametric modulation of the boson channel.  Here $\theta$ is the relative phase between the parametric drive and the coupling interaction in the first term. This general Hamiltonian can be realized in many physical systems including trapped ions~\cite{Ge2019, Ge2019b}, cavity optomechanics~\cite{Lemonde2016}, and superconducting circuit quantum electrodynamics (QED)  or atom-cavity systems~\cite{Qin2018,Chen2019}. 

For interactions between trapped-ion qubits, the $\hat{a}$ and $\hat{a}^\dagger$ operators typically correspond to a normal mode of ion motion in the trap (whose excitations are phonons) and $\hat{s}_{i}=\hat{\sigma}_{j}^{i}$, where $\hat{\sigma}_j^{i}$ is a Pauli operator for the $i^\mathrm{th}$ ion with $j\in\{x,y,z\}$~\cite{Sorensen1999, Sackett2000,Milburn2000,Leibfried2003exp}.  Here $\delta$ is the detuning of the spin-motion coupling drive from the frequency $\omega$ of the phonon mode used to implement the interaction, and $\beta_{i}$ describes the participation of the $i^\mathrm{th}$ ion in the phonon mode.  As another example, one could also use Eq.~(\ref{eq:H_M}) to describe atoms or superconducting qubits coupled to a single electromagnetic field mode in a cavity.  There $\hat{a}$ and $\hat{a}^\dagger$ are the cavity mode operators, while $\hat{s}_{i} = \hat{\sigma}^{i}_{+}$, where $\hat{\sigma}_{+}^{i}$ is the effective spin-1/2 raising operator for the $i^\mathrm{th}$ atom in the cavity. The $\beta_{i}$ then describe the relative atom-cavity coupling strengths, and $\delta$ is the atom-cavity detuning. When $g=0$, $\hat{H}_{M}$ is equivalent to the Tavis-Cummings Hamiltonian~\cite{Tavis1968}. 

\begin{figure}
\includegraphics[width=0.98\columnwidth]{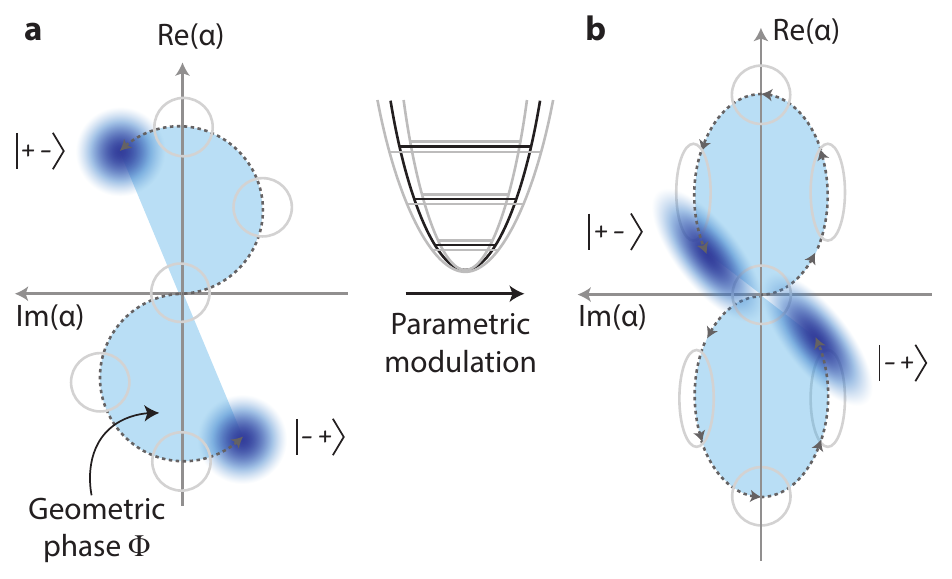}
\centering
\caption{\textbf{Phase space illustration of boson-mediated interactions.} We show the dimensionless complex harmonic oscillator amplitude $\alpha$ and corresponding schematic quasi-probability distributions\cite{Gerry2005} (shown in dark blue, with past positions outlined in grey) for two qubits ($\hat{S}=\hat{\sigma}_{x}^{1}-\hat{\sigma}_{x}^{2}$) assumed to be initialized at the origin of phase space. In both panels, the interaction duration is the same. \textbf{a}, Without parametric modulation ($g=0$), the wave packets associated with the $\ket{+-}$ and $\ket{-+}$ spin states follow circular trajectories in a frame rotating at $\omega+\delta$. The enclosed area (shown in light blue) is equal to the acquired geometric phase $\Phi$. \textbf{b}, With parametric modulation at $2\omega+2\delta$, the trajectories (again viewed in a frame rotating at $\omega+\delta$) become elliptical and the motional wave packets are alternately squeezed and anti-squeezed. The parametric modulation results in increased geometric phase being acquired per unit time. }
\label{fig:Squeezed_gate_ps}
\end{figure}

As a representative case, we examine the dynamics in a trapped-ion system without parametric modulation ($g=0$). We consider two trapped-ion qubits (with single-qubit $\hat{\sigma}_z$ eigenstates $\ket{\uparrow}$ and $\ket{\downarrow}$) coupled through a shared out-of-phase mode of motion such that $\hat{S}=\hat{\sigma}_{x}^{1}-\hat{\sigma}_{x}^{2}$ (note $\hat{S}=\hat{S}^\dagger$). Applying $\hat{H}_{M}$ will result in spin-dependent displacements in the phase space of the motional mode \cite{Milburn2000,Leibfried2003exp}. The two-qubit spin states $\ket{++}$ and $\ket{--}$, where $\ket{\pm}\equiv\frac{1}{\sqrt{2}}(\ket{\uparrow}\pm\ket{\downarrow})$, are not displaced, whereas states $\ket{+-}$ and $\ket{-+}$ will traverse circular trajectories in phase space (see Fig.~\ref{fig:Squeezed_gate_ps}a), each acquiring a state-dependent geometric phase $\Phi$ equal to the area enclosed by its trajectory~\cite{Milburn2000,Leibfried2003exp}. Applying $\hat{H}_{M}$ for a duration $\tau=2\pi/\delta$ returns the harmonic oscillator to its initial state after a single phase-space loop and disentangles it from the spin states. This results in the propagator~\cite{Molmer1999b}

\begin{equation}
\hat{U}=\exp \left( i\frac{\Phi}{4}\hat{S}^{2}\right),
\label{eq:USS}
\end{equation}

\noindent
which generates an effective spin-spin interaction, where $\Phi=2\pi(\Omega_0/\delta)^2$.  When $\delta=2\Omega_0$ and $\Phi=\pi/2$ this results in the maximally entangled state: $\hat{U}\ket{\downarrow \downarrow} = {\frac{1}{\sqrt{2}}(\ket{\downarrow \downarrow}+i\ket{\uparrow \uparrow})}$. If we include parametric modulation ($g\neq0$), the dynamics can be elucidated by making the normal mode transformation~\cite{Bogoliubov1958} $\hat{b}\equiv\hat{a}\cosh r -\hat{a}^{\dag}e^{i\theta} \sinh r $, where $r= \frac{1}{4}\ln\left[(\delta+g)/(\delta-g)\right]$. When $\theta=0$, this gives 

\begin{eqnarray}
\hat{H}_{M}&=&\frac{\hbar G\Omega_{0}}{2}(\hat{S}^{\dagger}\hat{b}+\hat{S}\hat{b}^{\dagger})-
\hbar\delta'\hat{b}^{\dagger}\hat{b}, 
\label{eq:H_PM}
\end{eqnarray}

\noindent
where $G=\left[(\delta+g)/(\delta-g)\right]^{1/4}=e^{r}$ and $\delta'=\sqrt{\delta^{2}-g^{2}}$, with the requirement that $|\delta|>|g|$. The values of $\delta$ and $\tau$ depend on $\Omega_0$ and $g$, and are determined by numerically solving a system of nonlinear equations (see Methods). The transformed Hamiltonian has the mathematical form of a boson-mediated interaction without parametric driving, but the interaction strength has been increased by a factor of $G$. Similarly, we can derive the propagator as in Eq.~\ref{eq:USS}, with the duration to acquire a given geometric phase $\Phi$ reduced by the same factor of $G$. The choice of $\theta=0$ gives the maximum amplification of the interaction strength; other values of $\theta$ provide less amplification, or even de-amplification (see Methods). The parametric modulation causes the $\ket{+-}$ and $\ket{-+}$ states to traverse elliptical, rather than circular, trajectories in phase space (Fig.~\ref{fig:Squeezed_gate_ps}b). Physically, the parametric modulation alternately squeezes and anti-squeezes the oscillator wave packets as they follow their elliptical trajectories, resulting in amplification of the spin-dependent displacements~\cite{Burd2019,Ge2019b}. For the case where $\hat{S}\neq\hat{S}^\dagger$, the result of parametric modulation depends on the details of $\hat{S}$.  For example, in cavity or circuit QED, the increase in interaction strength is given by $\cosh(r)$, and the amplification is independent of $\theta$, as shown in Ref.~\onlinecite{Qin2018}.

In our experiment, we amplify boson-mediated interactions between two trapped $^{25}$Mg$^{+}$ ion hyperfine qubits. The ions are held ${\simeq}\,30\,\mu$m above a linear surface-electrode radio-frequency trap~\cite{Seidelin2006, Srinivas2019, Burd2019} operated at 15\,K.  The harmonic oscillator mode is an out-of-phase radial motional mode with frequency $\omega \simeq 2\pi\times 5.9\,$MHz. We use qubit states $\ket{\downarrow}\equiv\ket{F=3, m_F=1}$ and $\ket{\uparrow}\equiv\ket{F=2, m_F=1}$ within the $^{2}S_{1/2}$ electronic ground state hyperfine manifold, where $F$ is the total angular momentum and $m_F$ is its projection along the quantization axis defined by a 21.3$\,$mT magnetic field. At this field strength, the qubit transition frequency $\omega_{0}\simeq 2\pi\times 1.686\,$GHz is insensitive to magnetic field fluctuations to first order, resulting in a qubit coherence time longer than 200\,ms. Global qubit rotations and coherent population transfer between hyperfine states as required for state preparation and readout are performed by applying resonant microwave pulses to trap electrodes. 

In each experiment, the ions are initialized in the electronic ground state $\ket{\downarrow\downarrow}$, and close to the motional ground state (mean occuption $\bar{n}\approx0.3$ in the phonon mode used to mediate interactions), with optical pumping, resolved-sideband laser cooling~\cite{Monroe1995}, and microwave pulses. Qubit readout is accomplished by transferring the population in $\ket{\downarrow}$ to $^{2}S_{1/2}\ket{F=3,m_F=3}$, applying a laser resonant with the $^{2}S_{1/2}\ket{F=3,m_F=3}\,\,\leftrightarrow\,\, ^{2}P_{3/2}\ket{F=4,m_F=4}$ cycling transition, and detecting state-dependent ion fluorescence. Coupling between qubits and phonons associated with the shared motional mode is implemented using the M{\o}lmer-S{\o}rensen (MS) interaction~\cite{Sorensen1999, Molmer1999b}. We implement the MS interaction using oscillating near-field magnetic field gradients at $\omega_0\pm(\omega+\delta)$, generated by currents in the trap electrodes (see Methods)~\cite{Ospelkaus2008, Ospelkaus2011}. We measure ${\Omega_{0}/2\pi=1.46(1)}$\,kHz, corresponding to a nominal single-loop MS gate duration ($\tau=\pi/\Omega_{0}$) of 342(3)$\,\mu$s.   Generating the MS interaction in this way enables straightforward phase synchronization with the parametric modulation at $2\omega+2\delta$, which is implemented by applying an oscillating potential directly to the rf trapping electrodes as described in Ref.~\citenum{Burd2019}.  

\begin{figure}
\includegraphics[width=0.98\columnwidth]{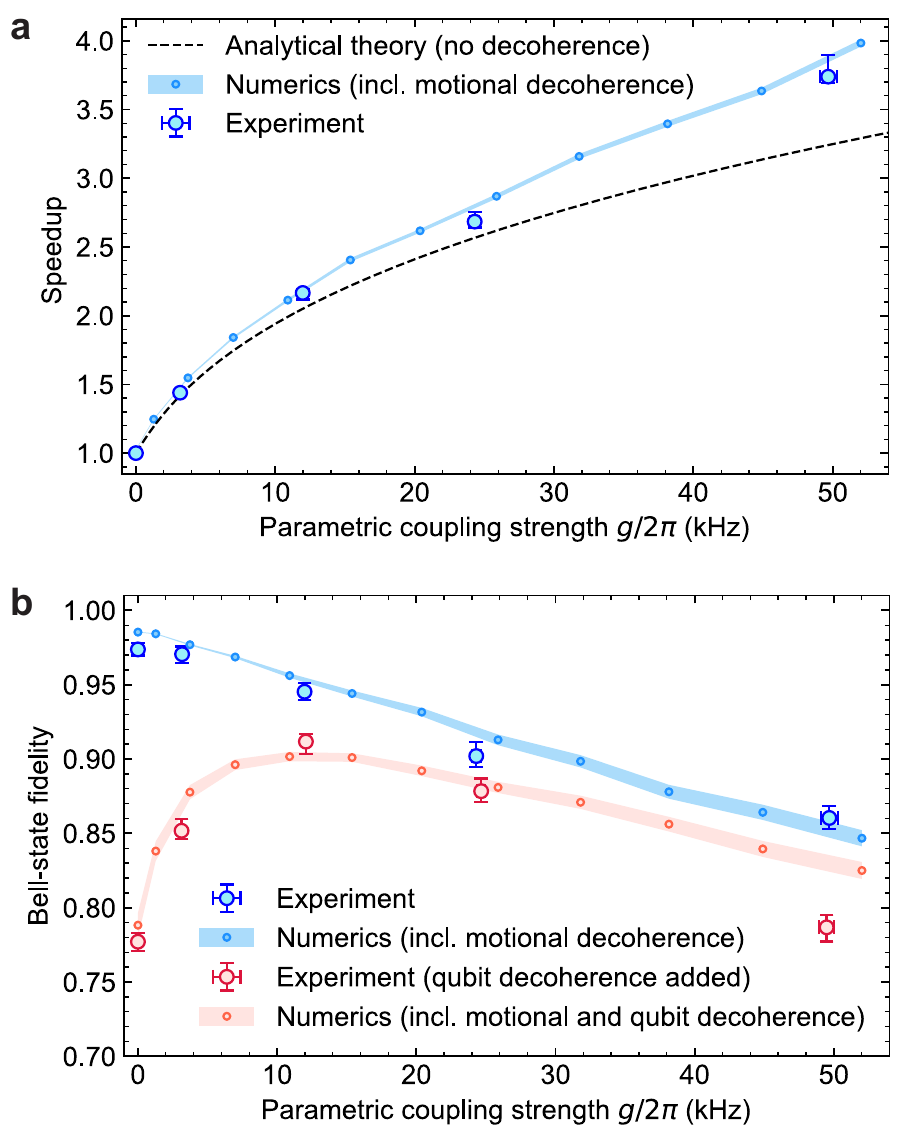}
\centering
\caption{\textbf{Bell-state fidelities and speedup.}  \textbf{a}, Speedup $t_{0}/\tilde{t}_{I,est}$ as a function of $g$. The dashed line is the prediction from analytical theory without motional decoherence.  The small points are the results of numerical simulations including motional decoherence, with the colored bands representing the corresponding 68 \% confidence intervals, linearly interpolated between simulated points~\cite{Ge2019}. \textbf{b}, Bell-state fidelities with and without added qubit dephasing. We plot $\tilde{F}_{exp}$, the maximum measured fidelity for each value of $g$, determined by scanning over a 2D grid of interaction times $t_I$ and detunings $\delta'$. Red circles (blue circles) indicate data taken with (without) added qubit dephasing noise (see text and Methods section). Small points show numerical simulations of the maximum fidelity (varying $t_I$ and $\delta'$) as a function of $g$, including motional decoherence, and either with (red) or without (blue) added qubit decoherence.  The colored bands represent 68 \% confidence intervals on the numerically simulated values, linearly interpolated between simulated points.  In both panels, experimental error bars indicate $68\,\%$ confidence intervals.}
\label{fig:fidelities}
\end{figure}

 To quantify the enhancement in the interaction strength due to parametric amplification, we find, for a given parametric coupling strength $g$, the optimum interaction duration for preparing the Bell state $\ket{\psi_B}=\frac{1}{\sqrt{2}}(\ket{\downarrow\downarrow}+i \ket{\uparrow\uparrow})$ from the initial state $\ket{\downarrow\downarrow}$. In the absence of decoherence, this corresponds to acquiring a geometric phase of $\Phi=\pi/2$, using a single phase space loop. To determine the optimal interaction duration we perform measurements for different values of the interaction duration $t_I$ and detuning $\delta^{\prime}$ at each $g$, and use the fidelity $F(g,t_I,\delta^{\prime}) = \bra{\psi_B}\hat{\rho}(g, t_I, \delta^{\prime})\ket{\psi_B}$ of the prepared state as a success metric, where $\hat{\rho}$ is the density matrix of the prepared state~\cite{Keith2018}.  For each $g$, we perform a 2D quadratic fit to the measured fidelity values versus $t_{I}$ and $\delta^{\prime}$ and use the $t_I$ value of the fit function maximum as the estimated optimal interaction duration $\tilde{t}_{I, est}$ (see Methods).  With $t_0\equiv\tilde{t}_{I,est}(g=0)=331(1)\,\mu$s denoting the estimated optimal interaction duration without parametric amplification, we plot the measured gate speedup $t_0/\tilde{t}_{I,est}$ versus $g$ in Fig.~\ref{fig:fidelities}a.  The values of $\tilde{t}_{I,est}$ tend to be shorter than predicted by analytical theory without decoherence, thus giving a higher-than-expected measured speedup.  This effect arises from a tradeoff between fidelity reductions from amplified motional decoherence (which penalize longer $t_I$) and from incorrect geometric phase acquisition or failure to close the phase space loop (which penalize $t_I$ either shorter or longer than the optimal duration without decoherence).  The measured $\tilde{t}_{I,est}$ values agree quantitatively with numerical simulations incorporating motional heating and dephasing mechanisms. For the strongest parametric coupling of $g/2\pi=49.7(6)$~kHz, we measure a speedup factor of $3.74\substack{+0.16 \\ -0.04}$ experimentally. With this $g$ and the independently calibrated $\Omega_0$, we calculate a theoretical enhancement in the phonon-mediated interaction strength of $G=3.25(1)$ (see Methods).
 
 The amplified motional decoherence also causes the interaction fidelity to diminish as $g$ is increased.  We define the maximum experimentally measured fidelity $\tilde{F}_{exp}(g)$ for a given $g$ (scanning over $\delta^{\prime}$ and $t_I$, see Methods) and plot $\tilde{F}_{exp}$ for each $g$ in Fig.~\ref{fig:fidelities}b.  For a parametric coupling strength of $g/2\pi=49.7(6)$\,kHz, we measure $\tilde{F}_{exp} = 0.860(8)$ for $t_{I}=90\,\mu$s, compared to $\tilde{F}_{exp}= 0.974(4)$ without parametric amplification at $t_{I}=350\,\mu$s.  
 
\begin{figure}[h]
\includegraphics[width=0.98\columnwidth]{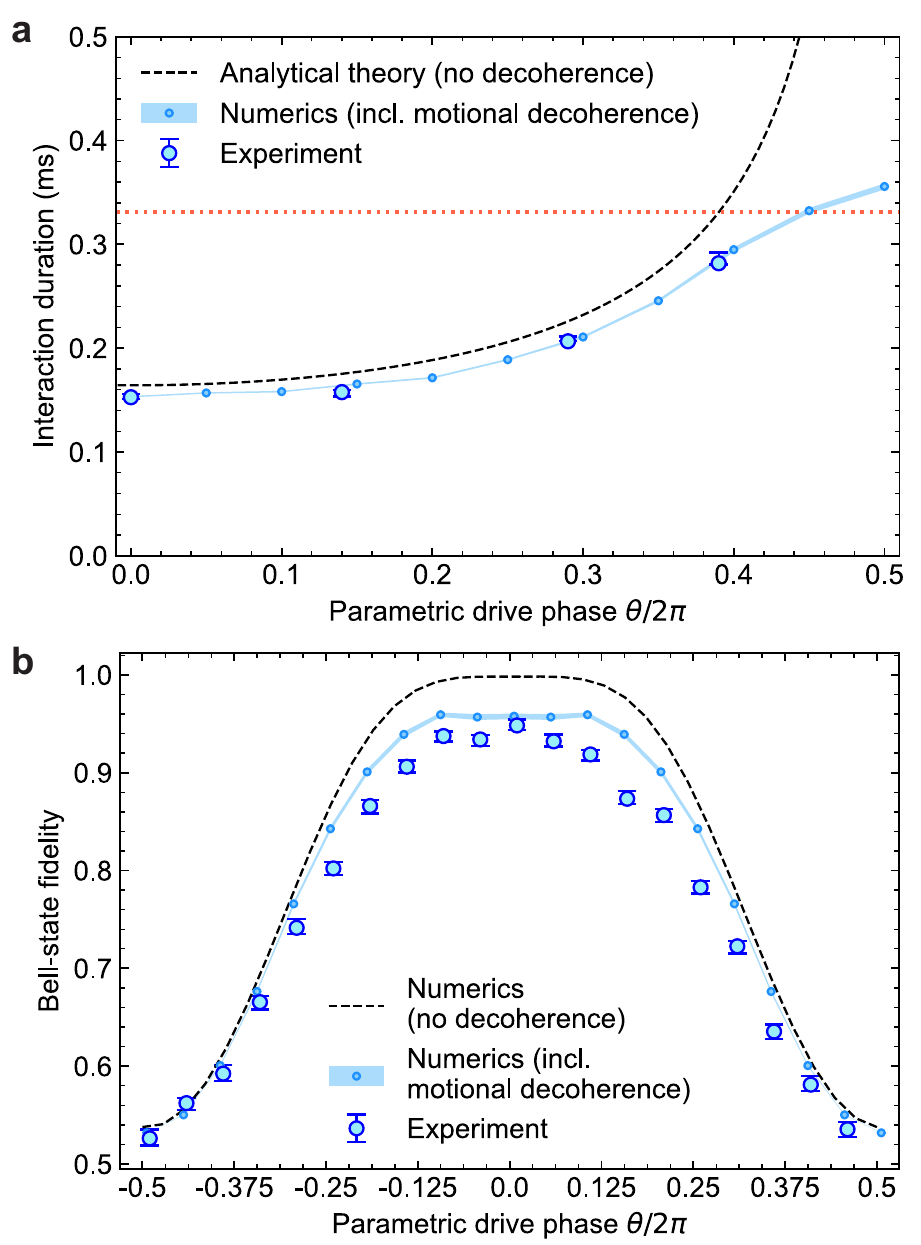}
\centering
\caption{\textbf{Phase dependence of amplification.}  \textbf{a}, Interaction duration $\tilde{t}_{I,est}$ as a function of the parametric drive phase $\theta$ for $g/2\pi=12.4(2)$\,kHz. The dotted red line indicates the duration $t_0$ without parametric amplification. The dashed black line is from analytical theory, without considering decoherence.  \textbf{b}, Fidelity as a function of $\theta$ for fixed $g$, $t_{I}$, and $\delta$, with $t_{I}$ and $\delta$ chosen to maximize fidelity at $\theta=0$ for $g/2\pi=12.4(2)$\,kHz. The dashed line is from numerical simulation, without considering decoherence.  In both panels, the small points are the result of numerical simulations including motional decoherence (see Methods), with the colored bands representing the 68 \% confidence intervals, linearly interpolated between simulated points.  In both panels, error bars on experimental data indicate 68\% confidence intervals.}
\label{fig:squeeze_phase}
\end{figure}

In our experiment, decoherence of the shared boson mode (ion motion) is the dominant source of gate infidelity.  However, decoherence mechanisms not arising from the shared boson mode (for example, off-resonant photon scattering from laser control fields~\cite{Ozeri2007,Uys2010}) are not amplified; in cases where these are the dominant source of infidelity, parametric amplification may improve the interaction fidelity by reducing the interaction duration~\cite{Ge2019}. As a proof-of-principle demonstration of operation in this regime, we introduce excess qubit dephasing by applying a current oscillating near $\omega_{0}$ to one of the trap electrodes. The detuned current gives rise to an ac Zeeman shift of the qubit frequency. The amplitude of the current is randomly changed every millisecond to give a Gaussian distribution (standard deviation of $2\pi\times 0.47(2)$\,kHz, see Methods) of qubit frequency shifts in time. In the presence of this dephasing, the maximum $\tilde{F}_{exp}$ without parametric modulation is 0.777(6), at $t_I=310\,\mu$s, whereas with parametric modulation ($g/2\pi=12.1(1)$\,kHz) we measure a maximum $\tilde{F}_{exp}$ of 0.912(7) at $t_I=160 \,\mu$s (Fig.~\ref{fig:fidelities}b red points).  Further increases in $g$ reduce the fidelity, due to errors from amplified motional decoherence.

In the experiments described above, the phase $\theta$ of the parametric drive with respect to the MS fields was set to give maximum amplification ($\theta=0$).  We can also vary $\theta$ to study the phase sensitivity of the amplification protocol. In Fig.~\ref{fig:squeeze_phase}a, we determine $\tilde{t}_{I,est}$ for different $\theta$ values and a fixed $g/2\pi= 12.4(2)$\,kHz. The parametric amplification process is theoretically third-order insensitive to small errors in the parametric drive phase~\cite{Ge2019}, and we measure no significant increase in $\tilde{t}_{I,est}$ up to at least $\theta/2\pi \approx 0.14$. Increasing $\theta$ further results in a longer $\tilde{t}_{I,est}$, in accordance with theoretical predictions. The increased $\tilde{t}_{I,est}$ is accompanied by a reduction in the maximum value of $\tilde{F}_{exp}$ over $t_I$ for each phase value, going from $0.945(6)$ to $0.738(8)$ as $\theta/2\pi$ changes from $0$ to $0.39$.  Simulations show that this fidelity loss is due to parametric amplification of motional decoherence. In Fig.~\ref{fig:squeeze_phase}b, we show the effect of varying $\theta$ for the same fixed $g$, without reoptimizing the gate parameters. While in our system the control fields used to induce boson-mediated interactions are at microwave frequencies and can therefore be readily phase-stabilized with respect to the parametric drive, applications with laser-based control fields may require stabilization of the laser optical phase at the ion positions relative to the parametric drive phase.  The fact that the interaction duration and fidelity are not first-order sensitive to this phase difference eases the requirements for the laser phase stability.

In summary, we have demonstrated parametric amplification of boson-mediated interactions between two trapped-ion qubits. Our method should increase entangling-gate fidelities in systems where the dominant sources of error result from qubit decoherence or from qubit errors induced by control fields, rather than decoherence of the bosonic degree of freedom that couples the qubits. Furthermore, the enhanced interaction strength afforded by parametric amplification could enable a reduction in the amount of laser or microwave power required in larger scale quantum information processors. Finally, we anticipate that parametric amplification will enable exploration of new parameter regimes in a variety of physical systems where boson-mediated interactions are essential. Possibilities include investigations of the dynamical Casimir effect~\cite{Qin2018}, photon-induced superconductivity~\cite{Babadi2017}, and enhanced spin squeezing in trapped ions~\cite{Ge2019} and neutral atoms~\cite{Qin2019,Groszkowski2020}.

\begin{acknowledgments}
We thank R. W. Simmonds, J. Schmidt, and L. J. Stephenson for a careful reading of the manuscript.  These experiments were performed using the ARTIQ control system. At the time the work was performed, S.C.B., R.S., H.M.K., and D.T.C.A. were Associates in the Professional Research Experience Program (PREP) operated jointly by NIST and the University of Colorado. This work was supported the NIST Quantum Information Program. S.C.B. carried out the experiments with assistance from D.H.S., R.S., H.M.K., and D.T.C.A., based on protocols developed by W.G. and J.J.B.; D.T.C.A., D.H.S., R.S., S.C.B., and H.M.K. built and maintained the apparatus; S.C.B., H.M.K., W.G., and D.H.S. analyzed the data and performed simulations; S.C.B. wrote the manuscript with input from all authors; and D.H.S. supervised the work, with support from J.J.B., D.T.C.A., D.L., A.C.W., and D.J.W.
\end{acknowledgments}

\section{Methods}

\subsection{Derivation of Eq.~\ref{eq:H_M} for combined M{\o}lmer-S{\o}rensen interaction and parametric modulation}

Here we show explicitly that the combined M{\o}lmer-S{\o}rensen (MS) and parametric interactions can be described by Eq. \ref{eq:H_M}. We consider two co-trapped atomic ions with internal qubit states $\ket{\downarrow}$ and $\ket{\uparrow}$ with energy separation $\hbar\omega_{0}$. Interactions between the qubits are mediated by a shared out-of-phase motional mode with frequency $\omega$. Without driving fields, the lab-frame Hamiltonian for the system is given by~\cite{Wineland1998} 
\begin{equation}
\hat{H}_{0}=\hbar\omega\hat{a}^{\dagger}\hat{a}+\frac{\hbar\omega_{0}}{2}(\hat{\sigma}_{z}^{1}+\hat{\sigma}_{z}^{2}),
\end{equation}

\noindent
where $\hat{\sigma}_{z}^{i}=\ket{\uparrow}\bra{\uparrow}-\ket{\downarrow}\bra{\downarrow}$ is a Pauli operator for ion $i$ and $\hat{a}$ ($\hat{a}^{\dagger}$) is the annihilation (creation) operator for the phonon mode. The MS interaction is implemented by simultaneously applying red (RSB) and blue (BSB) sideband interactions\cite{Sorensen1999}

\begin{eqnarray}
\hat{H}_{MS}&=&\hat{H}_{BSB}+\hat{H}_{RSB}\nonumber\\
&=&\hbar\Omega_{0}(\hat{\sigma}_{+}^{1}-\hat{\sigma}_{+}^{2})\hat{a}^{\dagger}\cos(\omega_{BSB}t)\nonumber\\
&&+\hbar\Omega_{0}(\hat{\sigma}_{+}^{1}-\hat{\sigma}_{+}^{2})\hat{a}\cos(\omega_{RSB}t)+\text{h.c.},
\end{eqnarray}

\noindent
where h.c.~is the Hermitian conjugate, $\Omega_{0}$ characterizes the qubit-motion coupling strength, and $\omega_{BSB}$ and $\omega_{RSB}$ are the frequencies of the sideband drives.  If the sidebands are symmetrically detuned from $\omega_{0}$ such that $\omega_{BSB}=\omega_{0}+\omega+\delta$, and $\omega_{RSB}=\omega_{0}-\omega-\delta$, we can transform into an interaction picture with respect to $\hat{H}_{0}$ to obtain~\cite{Sorensen1999}

\begin{eqnarray}
\hat{H}_{MS_{I}}&=&e^{i\hat{H}_{0}t/\hbar}\hat{H}_{MS}e^{-i\hat{H}_{0}t/\hbar}\nonumber\\
&=&\frac{\hbar\Omega_{0}}{2}(\hat{a}e^{+i\delta t }+\hat{a}^{\dagger}e^{-i\delta t})(\hat{\sigma}_{x}^{1}-\hat{\sigma}_{x}^{2}) ,
\end{eqnarray}
\noindent
where we have made a rotating wave approximation and dropped terms oscillating near $2\omega_{0}$.   The minus sign between the Pauli operators arises because the shared motional mode is an out-of-phase mode ($\beta_1=-\beta_2=1$, recalling the definition $\hat{S}\equiv\sum_{i}\beta_i\hat{s}_{i}$, where here $\hat{s}_{i} = \hat{\sigma}_x^i$).

We now consider the parametric drive. In the lab frame, modulation of the confining potential of a trapped-ion mechanical oscillator at frequency $\omega_{p}$ results in the Hamiltonian~\cite{Heinzen1990}

\begin{equation}
\label{eq:ham_HP}
\hat{H}_{P}=\hbar g(\hat{a}+\hat{a}^{\dagger})^{2}\cos(\omega_{P}t+\theta),
\end{equation}

\noindent
where $g$ is the parametric coupling strength and $\theta$ is the phase of the parametric drive relative to the MS interaction fields. If $\omega_P=2\omega+2\delta$, then the parametric drive Hamiltonian, in the interaction picture with respect to $\hat{H}_{0}$ and after making a rotating wave approximation, becomes

\begin{equation}
\hat{H}_{P_{I}}=\frac{\hbar g}{2}(\hat{a}^{2}e^{2i\delta t+i\theta}+\hat{a}^{\dagger2}e^{-2i\delta t- i\theta}).
\end{equation}

\noindent
Applying the MS fields and parametric drive simultaneously yields:

\begin{eqnarray}
\label{eq:ham_int}
\hat{H}_{I}&=&\hat{H}_{MS_{I}}+\hat{H}_{P_{I}}\nonumber\\
&=&\frac{\hbar\Omega_{0}}{2}(\hat{a}e^{i\delta t }+\hat{a}^{\dagger}e^{-i\delta t})(\hat{\sigma}_{x}^{1}-\hat{\sigma}_{x}^{2})\nonumber\\
&&+\frac{\hbar g}{2}(\hat{a}^{2}e^{2i\delta t+i\theta}+\hat{a}^{\dagger2}e^{-2i\delta t- i\theta}).
\end{eqnarray}

\noindent
If we transform Eq. \ref{eq:ham_int} into the interaction picture with respect to $\hat{H}_{1}=\hbar\delta \hat{a}^{\dagger}\hat{a}$, the time dependence can be eliminated, giving

\begin{eqnarray}
\label{eq:ham_ge}
e^{i\frac{\hat{H}_{1}t}{\hbar}}\hat{H}_{I}e^{-i\frac{\hat{H}_{1}t}{\hbar}}&=&\frac{\hbar\Omega_{0}}{2}(\hat{a}+\hat{a}^{\dagger})(\hat{\sigma}_{x}^{1}-\hat{\sigma}_{x}^{2}) \nonumber\\
&&+\frac{\hbar g}{2}(\hat{a}^{2}e^{ i\theta}+\hat{a}^{\dagger 2}e^{ -i\theta})-\hbar\delta \hat{a}^{\dagger}\hat{a},\nonumber \\
\end{eqnarray}

\noindent
which is equivalent to Eq. \ref{eq:H_M} with $\hat{S}=\hat{\sigma}_{x}^{1}-\hat{\sigma}_{x}^{2}$, using the fact that $\hat{S}=\hat{S}^\dagger$ by this definition.

\subsection{Phase-dependence of parametric amplification}

Applying the normal mode (Bogoliubov) transformation \cite{Bogoliubov1958} {$\hat{b}\equiv\hat{a}\cosh r -\hat{a}^{\dag}e^{i\theta} \sinh r $} to Eq. \ref{eq:H_M} gives the expression

\begin{eqnarray}
\hat{H}_{M}&=&\frac{\hbar\Omega_{0}}{2}\left( \hat{S}^{\dagger}\left[\hat{b}\cosh r+\hat{b}^{\dagger}\,e^{i\theta}\sinh r\right] \right. \nonumber\\
&&\left.+\hat{S}\left[\hat{b}^{\dagger}\cosh r+\hat{b}\,e^{-i\theta}\sinh r\right]      \right)-\hbar\delta'\hat{b}^{\dagger}\hat{b},
\label{eq:trMSPM}
\end{eqnarray}

\noindent
where $r= \frac{1}{4}\ln\left[(\delta+g)/(\delta-g)\right]$ and  $\delta'=\sqrt{\delta^{2}-g^{2}}$, with the requirement that $|\delta|>|g|$ and $\delta,g\geq0$.  For cases where $\hat{S}=\hat{S}^{\dagger}$,  we have

\begin{equation}
\hat{H}_{M}=\frac{\hbar\Omega_{0}}{2} \hat{S}\left(\hat{b} f(r,\theta) +\hat{b}^{\dagger}f^{*}(r,\theta)     \right)-\hbar\delta'\hat{b}^{\dagger}\hat{b},
\label{eq:trMSPM}
\end{equation}

\noindent
where $f(r,\theta)=\cosh r+e^{i\theta}\sinh r $. The interaction strength $\Omega_{0}$ is modified by a factor $|f(r,\theta)|$ (the phase of $f$ can be absorbed into the $\hat{b}$ operator), which is given by

\begin{equation}
|f(r,\theta)|=\sqrt{\cosh 2r+\cos \theta \sinh 2r }.
\label{eq:Modf}
\end{equation}

\noindent
Maximum amplification occurs when $\theta=0$, when $|f(r,0)|=e^{r}$ and the interaction strength becomes $\Omega_{0}e^{r}$. If $\theta=\pi$, the interaction strength is maximally suppressed, with $|f(r,\pi)|=e^{-r}$ and $\Omega_{0}\rightarrow \Omega_{0}e^{-r}$.  Equation~\ref{eq:Modf} can be reparameterized in terms of $\delta$ and $g$ to give:

\begin{equation}
|f(\delta, g,\theta)|=\left|\frac{\delta+g\cos(\theta)}{\sqrt{\delta^{2}-g^{2}}}    \right|^{1/2},
\label{eq:fdeltagtheta}
\end{equation}      

\noindent
 for $\delta>0$.  The above analysis also holds if $\delta<0$ (we always presume $g\geq 0$), provided that a phase shift of $\pi$ is added to $\theta$ as well.  

\subsection{Calculation of the gate time and detuning for parametrically amplified gates without decoherence}

The preparation of the Bell state $\ket{\psi_B}=\frac{1}{\sqrt{2}}(\ket{\downarrow\downarrow}+i \ket{\uparrow\uparrow})$ from the initial state $\ket{\downarrow\downarrow}$ requires the closure of an integer number of phase space loops and the accumulation of a geometric phase of $\Phi=\pi/2$. The closure of a single loop occurs when $\tau$ is given by:

\begin{equation}
\tau=\frac{2\pi}{\delta'}=\frac{2\pi}{\sqrt{\delta^{2}-g^{2}}}\, .
\label{eq:tgdelta}
\end{equation}

\noindent
The geometric phase acquired in that loop, $\Phi$, is:

\begin{equation}
\Phi=2\pi\left(\frac{\Omega_0|f(g,\delta,\theta)|}{\delta^{\prime}}\right)^{2}=2\pi\left(\frac{\Omega_0|f(g,\delta,\theta)|}{\sqrt{\delta^{2}-g^{2}}}\right)^{2}\, .
\label{eq:Phigdelta}
\end{equation}

\noindent
Given $g$, $\theta$, and $\Omega_0$, we can determine the correct values of $\tau$ and $\delta$ by numerically solving Eqs. \ref{eq:tgdelta} and \ref{eq:Phigdelta} simultaneously.  All values of $\tau$ from analytical theory shown in the text are calculated in this manner.  

\subsection{Calibration of the parametric drive strength $g$}

The electronics used to generate the parametric drive are described in detail in Ref.~\citenum{Burd2019} and consist of a direct digital synthesizer driving a resonant tank circuit coupled to the trap rf electrodes. Since a resonant circuit is used to couple the parametric drive to the trap electrodes, the parametric drive strength $g$ depends on the frequency $\omega_{P}$.
During entangling-gate experiments there are both slow drifts of the trap frequency $\omega$ and deliberate changes to the detuning $\delta$.  These result in small changes in $g$, since $\omega_{P}=2\omega+2\delta$. Since we measure $\omega$ at the beginning of every gate experiment, we can infer the value of $g$ for that experiment using an independently measured calibration function.

For given settings of the mode frequency $\omega$ and parametric drive amplitude, we measure $g$ using a similar method to that described in Ref.~\citenum{Burd2019}. First, we prepare the two-ion state $\ket{\uparrow\uparrow}$, with the motion cooled near the ground state ($\bar{n}\approx0.3$).  Next, we squeeze the motional state by applying the parametric drive on resonance (with $\omega_{P}=2\omega$) for a duration $t$. Parametric modulation ideally implements a squeezing operation~\cite{Heinzen1990,Walls1994} 

\begin{equation}
\ket{\xi}=\hat{Q}(\xi)\ket{0},
\end{equation}

\noindent where $\displaystyle \hat{Q}(\xi)=\textstyle\exp\left(\frac{1}{2}\left(\xi^{*}\hat{a}^{ 2}-\xi\hat{a}^{\dagger2}\right)\right)$ is the squeezing operator and $\ket{\xi}$ is a squeezed state characterized by the complex squeezing parameter $\xi=gte^{i\theta}$. The parametric coupling strength is given by $g=|\xi|/t$. The oscillator number state populations of the shared two-ion motional state can be inferred from the ions' qubit populations after applying a motion-adding sideband pulse of a variable duration and detecting the two-ion qubit populations. We can extract the value of $|\xi|$, and hence $g$, by fitting a numerical model that assumes the motion to be in a squeezed state $\ket{\xi}$ to the measured two-ion populations as a function of the duration of the sideband pulse, with only $|\xi|$ as a free parameter. Additional parameters used in the model are the sideband Rabi frequency for each ion, the ac Zeeman shift on each ion due to off-resonant magnetic fields associated with the sideband drive, and the initial thermal occupation $\bar{n}$, which are calibrated by fitting the model to data from a control experiment with the parametric drive amplitude set to zero. We repeat this experiment for various values of $\omega$ and fit a quadratic polynomial to the resulting data, obtaining a calibration function for $g$ as a function of $\omega$. All reported uncertainty values for $g$ are $68\%$ functional prediction intervals~\cite{tellinghuisen2001} based on this calibration function fit. Note that the prediction interval reflects our uncertainty in determining the underlying value of $g$, but not fluctuations of $g$ in time for a given $\omega$, which are significantly smaller. 

\subsection{Determination of $\tilde{t}_{I,est}$}

For each value of $g$ in the experiments described in Fig.~\ref{fig:fidelities}a, we measure the Bell-state fidelity over a grid of $\sim$\,25 $(t_{I},\delta)$ pairs. We then fit a 2D quadratic surface to the data to determine $\tilde{t}_{I,est}$. The fitting function and data for the $g=49.7(6)$\,kHz point in Fig.~\ref{fig:fidelities}a are given in the supplementary materials. The same method is used to determine $\tilde{t}_{I,est}$ for the data points shown in Fig.~\ref{fig:squeeze_phase}a. The vertical error bars shown in Fig.~\ref{fig:fidelities}a and Fig.~\ref{fig:squeeze_phase}a are the 68\% confidence intervals of the values of  $t_0/\tilde{t}_{I,est}$ and $\tilde{t}_{I,est}$, respectively, obtained using bootstrapping~\cite{Efron1994}.  For each data point plotted, we generate 5,000 nonparametrically resampled data sets, accounting for uncertainty both in the calibration of $g$ and in the estimated fidelities, and determine the central 68 \% confidence intervals in $t_0/\tilde{t}_{I,est}$ or $\tilde{t}_{I,est}$ from the distribution of the corresponding fitted values across all the resampled data sets.  Further details can be found in the Supplementary Material.  

\subsection{Excess qubit dephasing noise}

The $\ket{\downarrow}\leftrightarrow\ket{\uparrow}$ transition frequency is first-order insensitive to magnetic field fluctuations (a so-called ``clock" transition).  We therefore induce excess qubit frequency fluctuations, and thus dephasing, by applying a time-varying ac Zeeman shift $\delta_{ac}$.  We generate the ac Zeeman shift by applying a current oscillating near $\omega_0$ to generate an off-resonant oscillating magnetic field at the ion. The magnitude of $\delta_{ac}$ is proportional to the square of the applied current amplitude $I_{ac}$, which we verify experimentally by measuring the qubit frequency for different $I_{ac}$ values.  We change $I_{ac}$ between different randomly chosen amplitude values once every millisecond.  The time of this change is not synchronized with the rest of the experiment, such that over many experimental trials, the changes will occur at uniformly distributed random times with respect to the start of each trial.  The random current amplitudes are chosen to give qubit frequency fluctuations (which cause qubit dephasing) according to a Gaussian distribution with mean value $\bar{\delta}_{ac}/2\pi=4.59(1)\,$kHz and standard deviation $\sigma_{ac}/2\pi=0.47(2)\,$kHz.  

We characterize the effect of the applied qubit dephasing noise by performing Ramsey experiments as follows.  First, the qubits are initialized in the state $\ket{\downarrow \downarrow}$. A global carrier $\pi/2$ pulse is then applied to the qubits. Next, the fluctuating ac Zeeman shift is applied for duration $t_{R}$. A second carrier $\pi/2$ pulse completes the Ramsey sequence and the two-ion populations are measured. For a given distribution of applied ac Zeeman shift fluctuations, we measure the populations for various values of $t_{R}$. We fit these data to a numerical model of the expected populations, computed from an ensemble of simulated trials with random static qubit frequency shifts drawn from a Gaussian distribution. This fit enables us to determine the mean and standard deviation of the qubit frequency shifts due to the applied ac Zeeman shift noise. We cross-calibrate by measuring $\delta_{ac}$ as a function of $I_{ac}$ and calculating the mean and standard deviation of $\delta_{ac}$ based on the known distribution of $I_{ac}$ values applied. These two calibrations agree quantitatively.  

\subsection{Numerical Simulations}
For numerical simulations of the data obtained in the experiment, we include three types of motional decoherence processes: (1) motional dephasing due to the coupling to a phase-damping reservoir~\cite{Turchette2000}, (2) shot-to-shot fluctuations of the trap frequency, and (3) motional heating~\cite{Brownnutt2015}. Dephasing (1) can be modeled by the master equation for the system density matrix written as\cite{Turchette2000}
\begin{align}
\dot{\hat\rho}=\frac{\gamma}{2}\left[2\hat a^{\dagger}\hat a\hat\rho\hat a^{\dagger}\hat a-\left(\hat a^{\dagger}\hat a\right)^2\hat\rho-\hat\rho\left(\hat a^{\dagger}\hat a\right)^2\right],
\end{align}
where $\gamma$ is the dephasing rate. In our simulations, we take $\gamma$ to be a free parameter to fit the experimental data. The motional dephasing terms can be converted into a quantum stochastic equation for $\hat a$ given by
\begin{align}
\dot{\hat a}=-\gamma \hat a-i\sqrt{\gamma}\eta(t)\hat a,
\end{align}
where $\eta(t)$ describes white noise with the correlation $\braket{\eta(t+\tau)\eta(t)}=\gamma \delta(\tau)$. 
Shot-to-shot motional frequency fluctuations (2) are included by running the simulation many times, each time with a randomly chosen motional frequency.  The motional frequency values are chosen from a Gaussian distribution with standard deviation of $\sigma_{\delta}=2\pi\times 100$ Hz, consistent with the shot-to-shot variation in motional frequency seen experimentally. As we use an out-of-phase motional mode, motional heating (3) is strongly suppressed. We measure a heating rate of $\dot{\bar{n}}\sim1$ quanta/s on this mode, which has a relatively small effect on the gate fidelity. Heating is included phenomenologically using a method similar to the treatment of photon scattering~\cite{Ozeri2007}.

\subsubsection{Quadratic Hamiltonian interaction picture}
We perform the numerical simulations for the data in Figs.\,\,\ref{fig:fidelities} and \ref{fig:squeeze_phase}   in the interaction picture of the quadratic Hamiltonian $\hat{H}_Q(t)=\hat{H}_P-\hat{H}_1$, where $\hat{H}_P$ is the parametric drive Hamiltonian without the rotating-wave approximation in Eq. \ref{eq:ham_HP} and $\hat{H}_1=\hbar\delta\hat{a}^{\dagger}\hat{a}$. Since $\hat{H}_Q(t)$ is quadratic in $\hat a$ and $\hat a^{\dagger}$, the interaction-picture creation operator $\hat{a}_I^{\dagger}$ is given by
\begin{equation}
\hat{a}_I^{\dagger}=\hat{U}^{\dagger}_Q(t)\hat{a}^{\dagger}\hat{U}_Q(t)=u(t)\hat{a}^{\dagger}+v^{\ast}(t)\hat{a}\, ,
\end{equation}
\noindent
where $\hat{U}_Q(t)=\hat{\mathcal{T}}\exp\left(-i\int_{0}^t\hat{H}_Q(\tau)d\tau\right)$ and $\hat{\mathcal{T}}$ is the time-ordering operator. The equations to determine $u(t)$ and $v(t)$ are given by
\begin{align}
\dot{u}&=\left[-\gamma-i\left(\delta+\sqrt{\gamma}\eta(t)\right)\right]u+ig\left(e^{i\theta}+e^{-i2\omega_Pt-i\theta}\right)v,\nonumber\\
\dot{v}&=\left[-\gamma+i\left(\delta+\sqrt{\gamma}\eta(t)\right)\right]v-ig\left(e^{-i\theta}+e^{i2\omega_Pt+i\theta}\right)u.
\end{align}
In the interaction picture of $\hat{H}_Q(t)$, the boson-mediated interaction term of Eq. \ref{eq:H_M} becomes

\begin{eqnarray}
\hat{\mathcal{V}}(t)&=&\hat{U}^{\dagger}_{Q}(t)\frac{\hbar\Omega_{0}}{2}(\hat{S}^{\dagger}\hat{a}+\hat{S}\hat{a}^{\dagger}) \hat{U}_{Q}(t)\nonumber\\
&=&\hbar\frac{\Omega_0}{2}\left(\hat{S}^{\dagger}h^{\ast}(t)\hat{a}+\hat{S}h(t)\hat{a}^{\dagger}\right),
\end{eqnarray}
where $h(t)=u(t)+v(t)$ and $\hat{S}=\hat{\sigma}_{x}^{1}-\hat{\sigma}_{x}^{2}$. For example, without the parametric drive and the motional dephasing, $h(t)=e^{-i\delta t}$.  
As the interaction Hamiltonian $\hat{\mathcal{V}}(t)$ is only linear in $\hat a$ and $\hat a^{\dagger}$, the qubit-motion system can be written as
\begin{align}
\label{eq:whstate}
\ket{\Psi(t)}&=c_{++}(t)\ket{++}\ket{\alpha_{++}(t)}\nonumber\\
&+c_{+-}(t)\ket{+-}\ket{\alpha_{+-}(t)}\nonumber\\
&+c_{-+}(t)\ket{-+}\ket{\alpha_{-+}(t)}\nonumber\\
&+c_{--}(t)\ket{--}\ket{\alpha_{--}(t)},
\end{align}
where $c_{\pm\pm}(t)$ are time dependent coefficients for the qubit states $\ket{\pm\pm}$ and the $\ket{\alpha_{\pm\pm}(t)}$ are coherent states of motion defined by $\ket{\alpha_{\pm\pm}(t)}=\hat{\mathcal{D}}(\alpha_{\pm\pm}(t))\ket{0}$, where $\hat{\mathcal{D}}(\alpha)=\exp\left(\alpha\hat{a}^{\dagger}-\alpha^{\ast}\hat{a}\right)$ is the displacement operator\cite{Gerry2005}. In the eigenbasis of $\hat{S}$, we find $\alpha_{++}(t)=\alpha_{--}(t)=0$ and $\alpha_{+-}(t)=-\alpha_{-+}(t)=-i\int_0^t\Omega_0h(t^{\prime})dt^{\prime}$. With the initial qubit state $\ket{\downarrow\downarrow}$ and the target state $\ket{\psi_B}=\frac{1}{\sqrt{2}}\left(\ket{\downarrow\downarrow}+i\ket{\uparrow\uparrow}\right)$, the fidelity at the gate time $\tau$ can be evaluated as
\begin{align}
F(\tau)=\left|\frac{1}{2}-ic_{+-}(\tau)\exp\left(-|\alpha_{+-}(\tau)|^2/2\right)\right|^2.
\end{align}
We can then find the optimal fidelity with respect to the gate time $\tau$ for fixed $\delta$ and $g$. In the simulations of Figs. \ref{fig:fidelities} and \ref{fig:squeeze_phase}, we average the optimal fidelity over 600 simulations for each value of $g$. In each simulation, we draw a random motional frequency shift from a zero-mean Gaussian distribution with the standard deviation $\sigma_{\delta}=2\pi\times 100$ Hz. Every simulation has the same value of the motional dephasing rate $\gamma$.  We determine $\gamma\approx 2\pi\times5.2$ Hz by fitting the fidelity simulations to the data. 

\subsubsection{Zeeman shift fluctuations}
To show the robustness of the parametric amplification against fluctuations of the qubit frequency as shown in Fig. \ref{fig:fidelities}b (red data points and shaded band) in the main text, we introduce artificial qubit frequency fluctuations by applying a detuned oscillating current to one of the trap electrodes, as described above. The detuned current induces an ac Zeeman shift, whose magnitude is randomly varied by changing the current amplitude every millisecond to give a Gaussian distribution of qubit frequencies. Effectively, we add an additional term $\textstyle\hat H_z=\hbar\displaystyle\frac{\delta_{ac}}{2}(\hat{\sigma}_{z}^{1}+\hat{\sigma}_{z}^{2})$  to the system Hamiltonian, where $\delta_{ac}$ is drawn from a Gaussian distribution with mean  $\bar{\delta}_{ac}/2\pi=4.59(1)\,$kHz and standard deviation $\sigma_{ac}/2\pi=0.47$ kHz. In this case, the state cannot be written in the form of Eq. \eqref{eq:whstate} due to the fact that $\left[\hat{H}_z,\hat{S}\right]\ne0$ for $\hat{S}=\hat{\sigma}_{x}^{1}-\hat{\sigma}_{x}^{2}$. Instead, we write the state of the whole system as
\begin{align}
\ket{\Psi(t)}=\sum_{j=1}^3\sum_{k=0}^sC_{j,k}(t)\ket{\psi_j}\ket{k},
\end{align}
where $\ket{k}$ is the $k^\mathrm{th}$ Fock state of motion,  $\ket{\psi_1}=\frac{1}{\sqrt{2}}\left(\ket{++}+\ket{--}\right)$, $\ket{\psi_2}=\frac{1}{\sqrt{2}}\left(\ket{+-}+\ket{-+}\right)$, $\ket{\psi_3}=\frac{1}{\sqrt{2}}\left(\ket{+-}-\ket{-+}\right)$, and $\ket{\Psi(0)}=\frac{1}{\sqrt{2}}\left(\ket{\psi_1}-\ket{\psi_2}\right)\ket{0}$. The other eigenstate $\ket{\psi_0}=\frac{1}{\sqrt{2}}\left(\ket{++}-\ket{--}\right)$ of the two-qubit system is not involved because it is a dark state of the operators $\hat{H}_z$ and $\hat{S}$ and it is orthogonal to the initial state $\ket{\downarrow\downarrow}$. The set of equations is truncated at a certain motional Fock state $\ket{s}$ depending on the maximum displacement of the motional state $|\alpha_{\max}|\equiv\left|\int_0^{\tau/2}\Omega_0h(t^{\prime})dt^{\prime}\right|$. According to the Schr\"odinger equation under the Hamiltonian $\hat H_z+\hat {\mathcal{V}}(t)$, the differential equations of the coefficients are obtained as
\begin{align}
\dot{C}_{1,k}&=-i\delta_{ac}C_{2,k},\nonumber\\
\dot{C}_{2,k}&=-i\delta_{ac}C_{1,k}-i\Omega_0\left(C_{3,k+1}h^{\ast}\sqrt{k+1}+C_{3,k-1}h\sqrt{k}\right),\nonumber\\
\dot{C}_{3,k}&=-i\Omega_0\left(C_{2,k+1}h^{\ast}\sqrt{k+1}+C_{2,k-1}h\sqrt{k}\right).
\end{align}
The fidelity at the gate time is then given by
\begin{equation}
F(\tau)=\frac{1}{2}\left|C_{1,0}(\tau)-C_{2,0}(\tau)\right|^2.
\end{equation}
For example, at the maximum $g\sim2\pi \times50$ kHz, the maximum displacement is $|\alpha_{\max}|\sim 3$, and we verify that the truncation number $s=23$ is sufficient. 
For Fig.~\ref{fig:fidelities}b, we average the optimal fidelity over $600$ runs of simulations for each value of $g$, where both the motional frequency and the ac Zeeman shift are randomly chosen in each simulation.

\clearpage

\section{Supplementary materials}

\subsection{Fitting for $\tilde{t}_{I,est}$}

\begin{figure*}[tb]
\centering
\includegraphics[width=0.875\textwidth]{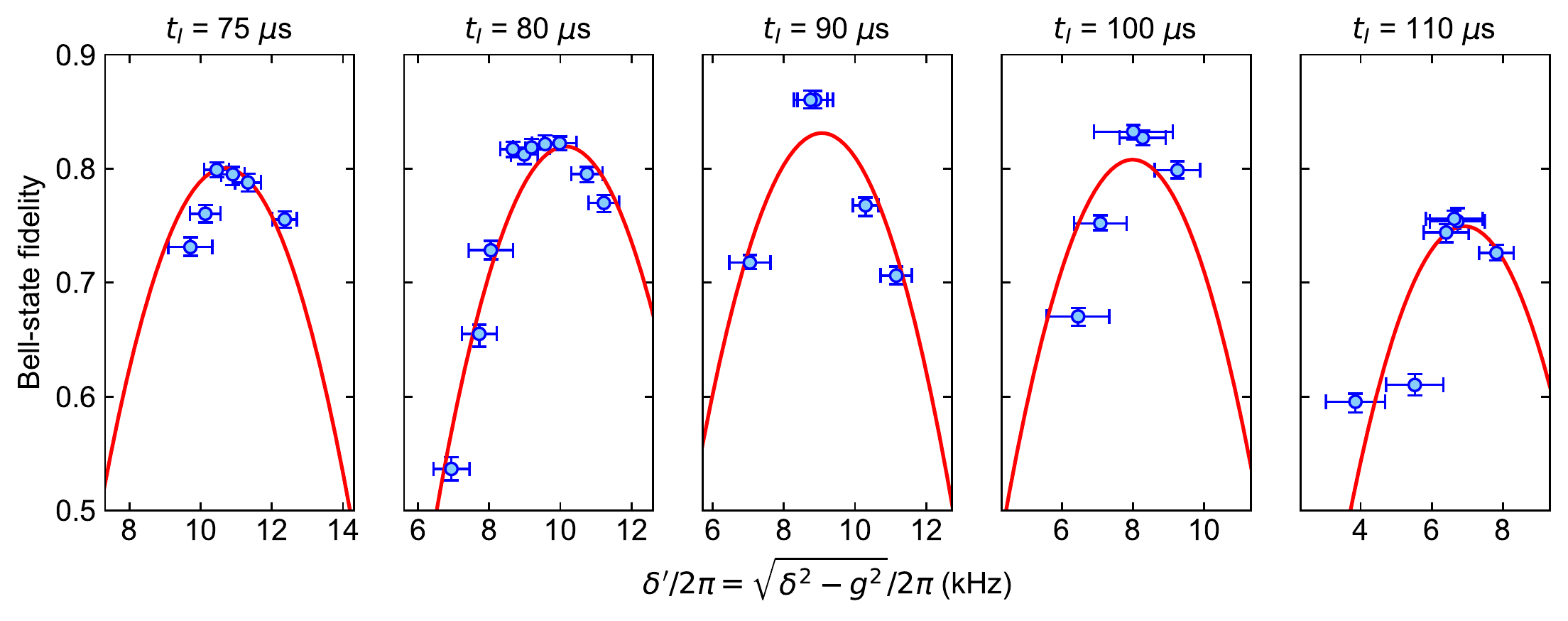}
\caption{Fidelity as a function of $\delta'/2\pi$ for various values of $t_{I}$ for a calibrated value of the parametric coupling strength of $g/2\pi=49.7(6)\,$kHz. Data points are fidelities obtained using the method described in Ref.~\citenum{Keith2018}. Vertical error bars indicate 68\,\% confidence intervals for the fidelity. Horizontal error bars indicate 68\,\% confidence intervals for $\delta'/2\pi$ calculated from error propagation of the measured uncertainties in $\delta$ and $g$ within a given experiment. Red curves are slices of the 2D quadratic fitting function at the corresponding interaction times.}
\label{fig:multi_plot}
\end{figure*}

For a given setting of the parametric drive amplitude $g$, we estimate the interaction duration required to maximize the two-qubit gate fidelity by fitting a 2D quadratic polynomial to the data. Specifically, we fit to a grid of experimentally measured fidelities at various $(t_{I},\delta)$ values to determine the quadratic constants $a_i$ in the function

\begin{eqnarray}
F(x,y)&=&  a_0 +a_1(x-a_4)^2+a_2(y-a_5)^2 \nonumber\\
&&+a_3(x-a_4)(y-a_5),
\end{eqnarray}

\noindent with $x\equiv t_{I}$, and $y \equiv\delta'$. Note that this requires converting the experimentally measured $\delta$ values into $\delta^\prime$ values, which requires knowledge of $g$.  Any calibration uncertainty in $g$ will propagate to the values of $\delta^\prime$.  The fitted value of the coefficient $a_4$ gives $\tilde{t}_{I,est}$. Data and slices of the 2D fit for the point at $g=2\pi\times 49.7(6)\,$kHz in Fig \ref{fig:fidelities}a are shown in Fig.~\ref{fig:multi_plot}. From the fit, $\tilde{t}_{I,est}=88\substack{+1 \\ -3}\,\mu$s. 

To determine the effect of the calibration uncertainty in $g$ on $\tilde{t}_{I,est}$ (arising from the propagation of this uncertainty into the $\delta^\prime$ values), as well as the uncertainty in the estimated fidelities, we use bootstrapping. For each setting of the parametric coupling strength, we generate $5,000$ synthetic data sets, each with the same number of $(t_I, \delta^\prime)$ data points as in the original data set. Each point in a new set has the original duration $t_{I}$, but has a detuning given by $\delta'=\sqrt{\delta^{2}-(g+\Delta g)^2}$, where $\Delta g$ is randomly selected from a zero-mean Gaussian distribution with standard deviation given by the calibration uncertainty in $g$ ($\sim 2\pi\times$ 600\,Hz for the data shown in Fig. \ref{fig:multi_plot}). Similarly, the fidelity values $F$ are given by $F+\Delta F$, where $\Delta F$ is randomly selected from a zero-mean Gaussian distribution with standard deviation given by the standard deviation of estimated fidelities from nonparametric bootstrapping of the raw gate data~\cite{Keith2018}.  For each synthetic data set we fit $F(x,y)$ to the data to determine $\tilde{t}_{I,est}$, and use the central 68 \% interval of the distribution of fitted values to determine the uncertainty in $\tilde{t}_{I,est}$.  

\begin{thebibliography}{10}
\expandafter\ifx\csname url\endcsname\relax
  \def\url#1{\texttt{#1}}\fi
\expandafter\ifx\csname urlprefix\endcsname\relax\def\urlprefix{URL }\fi
\providecommand{\bibinfo}[2]{#2}
\providecommand{\eprint}[2][]{\url{#2}}

\bibitem{Bruzewicz2019}
\bibinfo{author}{Bruzewicz, C.~D.}, \bibinfo{author}{Chiaverini, J.},
  \bibinfo{author}{McConnell, R.} \& \bibinfo{author}{Sage, J.~M.}
\newblock \bibinfo{title}{Trapped-ion quantum computing: Progress and
  challenges}.
\newblock \emph{\bibinfo{journal}{Appl. Phys. Rev.}}
  \textbf{\bibinfo{volume}{6}}, \bibinfo{pages}{021314} (\bibinfo{year}{2019}).

\bibitem{Blais2020}
\bibinfo{author}{Blais, A.}, \bibinfo{author}{Girvin, S.~M.} \&
  \bibinfo{author}{Oliver, W.~D.}
\newblock \bibinfo{title}{Quantum information processing and quantum optics
  with circuit quantum electrodynamics}.
\newblock \emph{\bibinfo{journal}{Nat. Phys.}} \textbf{\bibinfo{volume}{16}},
  \bibinfo{pages}{247–256} (\bibinfo{year}{2020}).

\bibitem{Georgescu2014}
\bibinfo{author}{Georgescu, I.~M.}, \bibinfo{author}{Ashhab, S.} \&
  \bibinfo{author}{Nori, F.}
\newblock \bibinfo{title}{Quantum simulation}.
\newblock \emph{\bibinfo{journal}{Rev. Mod. Phys.}}
  \textbf{\bibinfo{volume}{86}}, \bibinfo{pages}{153--185}
  (\bibinfo{year}{2014}).

\bibitem{Degen2017}
\bibinfo{author}{Degen, C.~L.}, \bibinfo{author}{Reinhard, F.} \&
  \bibinfo{author}{Cappellaro, P.}
\newblock \bibinfo{title}{Quantum sensing}.
\newblock \emph{\bibinfo{journal}{Rev. Mod. Phys.}}
  \textbf{\bibinfo{volume}{89}}, \bibinfo{pages}{035002}
  (\bibinfo{year}{2017}).

\bibitem{Pezze2018}
\bibinfo{author}{Pezz\`e, L.}, \bibinfo{author}{Smerzi, A.},
  \bibinfo{author}{Oberthaler, M.~K.}, \bibinfo{author}{Schmied, R.} \&
  \bibinfo{author}{Treutlein, P.}
\newblock \bibinfo{title}{Quantum metrology with nonclassical states of atomic
  ensembles}.
\newblock \emph{\bibinfo{journal}{Rev. Mod. Phys.}}
  \textbf{\bibinfo{volume}{90}}, \bibinfo{pages}{035005}
  (\bibinfo{year}{2018}).

\bibitem{Bloch2008}
\bibinfo{author}{Bloch, I.}, \bibinfo{author}{Dalibard, J.} \&
  \bibinfo{author}{Zwerger, W.}
\newblock \bibinfo{title}{Many-body physics with ultracold gases}.
\newblock \emph{\bibinfo{journal}{Rev. Mod. Phys.}}
  \textbf{\bibinfo{volume}{80}}, \bibinfo{pages}{885--964}
  (\bibinfo{year}{2008}).

\bibitem{Gerry2005}
\bibinfo{author}{Gerry, C.~C.} \& \bibinfo{author}{Knight, P.~L.}
\newblock \emph{\bibinfo{title}{Introductory {Q}uantum {O}ptics}}
  (\bibinfo{publisher}{Cambridge University Press},
  \bibinfo{address}{Cambridge, United Kingdom}, \bibinfo{year}{2005}).

\bibitem{Blais2004}
\bibinfo{author}{Blais, A.}, \bibinfo{author}{Huang, R.-S.},
  \bibinfo{author}{Wallraff, A.}, \bibinfo{author}{Girvin, S.~M.} \&
  \bibinfo{author}{Schoelkopf, R.~J.}
\newblock \bibinfo{title}{Cavity quantum electrodynamics for superconducting
  electrical circuits: An architecture for quantum computation}.
\newblock \emph{\bibinfo{journal}{Phys. Rev. A}} \textbf{\bibinfo{volume}{69}},
  \bibinfo{pages}{062320} (\bibinfo{year}{2004}).

\bibitem{Evans2018}
\bibinfo{author}{Evans, R.~E.} \emph{et~al.}
\newblock \bibinfo{title}{Photon-mediated interactions between quantum emitters
  in a diamond nanocavity}.
\newblock \emph{\bibinfo{journal}{Science}} \textbf{\bibinfo{volume}{362}},
  \bibinfo{pages}{662--665} (\bibinfo{year}{2018}).

\bibitem{Cirac1995}
\bibinfo{author}{Cirac, J.~I.} \& \bibinfo{author}{Zoller, P.}
\newblock \bibinfo{title}{Quantum computations with cold trapped ions}.
\newblock \emph{\bibinfo{journal}{Phys. Rev. Lett.}}
  \textbf{\bibinfo{volume}{74}}, \bibinfo{pages}{4091} (\bibinfo{year}{1995}).

\bibitem{Sorensen1999}
\bibinfo{author}{S{\o}rensen, A.} \& \bibinfo{author}{M{\o}lmer, K.}
\newblock \bibinfo{title}{Quantum computation with ions in thermal motion}.
\newblock \emph{\bibinfo{journal}{Phys. Rev. Lett.}}
  \textbf{\bibinfo{volume}{82}}, \bibinfo{pages}{1971} (\bibinfo{year}{1999}).

\bibitem{Milburn2000}
\bibinfo{author}{Milburn, G.~J.}, \bibinfo{author}{Schneider, S.} \&
  \bibinfo{author}{James, D. F.~V.}
\newblock \bibinfo{title}{Ion trap quantum computing with warm ions}.
\newblock \emph{\bibinfo{journal}{Fortschr. Phys.}}
  \textbf{\bibinfo{volume}{48}}, \bibinfo{pages}{801--810}
  (\bibinfo{year}{2000}).

\bibitem{Higginbotham2018}
\bibinfo{author}{Higginbotham, A.~P.} \emph{et~al.}
\newblock \bibinfo{title}{{Harnessing electro-optic correlations in an
  efficient mechanical converter}}.
\newblock \emph{\bibinfo{journal}{Nat. Phys.}} \textbf{\bibinfo{volume}{14}},
  \bibinfo{pages}{1038--1042} (\bibinfo{year}{2018}).

\bibitem{Lu2015}
\bibinfo{author}{L\"u, X.-Y.} \emph{et~al.}
\newblock \bibinfo{title}{Squeezed optomechanics with phase-matched
  amplification and dissipation}.
\newblock \emph{\bibinfo{journal}{Phys. Rev. Lett.}}
  \textbf{\bibinfo{volume}{114}}, \bibinfo{pages}{093602}
  (\bibinfo{year}{2015}).

\bibitem{Lemonde2016}
\bibinfo{author}{Lemonde, M.-A.}, \bibinfo{author}{Didier, N.} \&
  \bibinfo{author}{Clerk, A.~A.}
\newblock \bibinfo{title}{Enhanced nonlinear interactions in quantum
  optomechanics via mechanical amplification}.
\newblock \emph{\bibinfo{journal}{Nat. Commun.}} \textbf{\bibinfo{volume}{7}},
  \bibinfo{pages}{11338} (\bibinfo{year}{2016}).

\bibitem{Zeytino2017}
\bibinfo{author}{Zeytino\ifmmode~\breve{g}\else \u{g}\fi{}lu, S.},
  \bibinfo{author}{\ifmmode \dot{I}\else \.{I}\fi{}mamo\ifmmode~\breve{g}\else
  \u{g}\fi{}lu, A.} \& \bibinfo{author}{Huber, S.}
\newblock \bibinfo{title}{Engineering matter interactions using squeezed
  vacuum}.
\newblock \emph{\bibinfo{journal}{Phys. Rev. X}} \textbf{\bibinfo{volume}{7}},
  \bibinfo{pages}{021041} (\bibinfo{year}{2017}).

\bibitem{Qin2018}
\bibinfo{author}{Qin, W.} \emph{et~al.}
\newblock \bibinfo{title}{Exponentially enhanced light-matter interaction,
  cooperativities, and steady-state entanglement using parametric
  amplification}.
\newblock \emph{\bibinfo{journal}{Phys. Rev. Lett.}}
  \textbf{\bibinfo{volume}{120}}, \bibinfo{pages}{093601}
  (\bibinfo{year}{2018}).

\bibitem{Chen2019}
\bibinfo{author}{Chen, Y.-H.}, \bibinfo{author}{Qin, W.} \&
  \bibinfo{author}{Nori, F.}
\newblock \bibinfo{title}{Fast and high-fidelity generation of steady-state
  entanglement using pulse modulation and parametric amplification}.
\newblock \emph{\bibinfo{journal}{Phys. Rev. A}}
  \textbf{\bibinfo{volume}{100}}, \bibinfo{pages}{012339}
  (\bibinfo{year}{2019}).

\bibitem{Leroux2018}
\bibinfo{author}{Leroux, C.}, \bibinfo{author}{Govia, L. C.~G.} \&
  \bibinfo{author}{Clerk, A.~A.}
\newblock \bibinfo{title}{Enhancing cavity quantum electrodynamics via
  antisqueezing: Synthetic ultrastrong coupling}.
\newblock \emph{\bibinfo{journal}{Phys. Rev. Lett.}}
  \textbf{\bibinfo{volume}{120}}, \bibinfo{pages}{093602}
  (\bibinfo{year}{2018}).

\bibitem{Arenz2018}
\bibinfo{author}{Arenz, C.}, \bibinfo{author}{Bondar, D.~I.},
  \bibinfo{author}{Burgarth, D.}, \bibinfo{author}{Cormick, C.} \&
  \bibinfo{author}{Rabitz, H.}
\newblock \bibinfo{title}{Amplification of quadratic {H}amiltonians}.
\newblock \emph{\bibinfo{journal}{{Quantum}}} \textbf{\bibinfo{volume}{4}},
  \bibinfo{pages}{271} (\bibinfo{year}{2020}).

\bibitem{Ge2019}
\bibinfo{author}{Ge, W.} \emph{et~al.}
\newblock \bibinfo{title}{Trapped ion quantum information processing with
  squeezed phonons}.
\newblock \emph{\bibinfo{journal}{Phys. Rev. Lett.}}
  \textbf{\bibinfo{volume}{122}}, \bibinfo{pages}{030501}
  (\bibinfo{year}{2019}).

\bibitem{Ge2019b}
\bibinfo{author}{Ge, W.} \emph{et~al.}
\newblock \bibinfo{title}{Stroboscopic approach to trapped-ion quantum
  information processing with squeezed phonons}.
\newblock \emph{\bibinfo{journal}{Phys. Rev. A}}
  \textbf{\bibinfo{volume}{100}}, \bibinfo{pages}{043417}
  (\bibinfo{year}{2019}).

\bibitem{Groszkowski2020}
\bibinfo{author}{Groszkowski, P.}, \bibinfo{author}{Lau, H.-K.},
  \bibinfo{author}{Leroux, C.}, \bibinfo{author}{Govia, L. C.~G.} \&
  \bibinfo{author}{Clerk, A.~A.}
\newblock \bibinfo{title}{Heisenberg-limited spin-squeezing via bosonic
  parametric driving}.
\newblock \emph{\bibinfo{journal}{\emph{Preprint at
  https://arxiv.org/abs/2003.03345v1}}}  (\bibinfo{year}{2020}).

\bibitem{Li2020}
\bibinfo{author}{Li, P.-B.}, \bibinfo{author}{Zhou, Y.}, \bibinfo{author}{Gao,
  W.-B.} \& \bibinfo{author}{Nori, F.}
\newblock \bibinfo{title}{Enhancing spin-phonon and spin-spin interactions
  using linear resources in a hybrid quantum system}.
\newblock \emph{\bibinfo{journal}{\emph{Preprint at
  https://arxiv.org/abs/2003.07151}}}  (\bibinfo{year}{2020}).

\bibitem{Ballance2016}
\bibinfo{author}{Ballance, C.~J.}, \bibinfo{author}{Harty, T.~P.},
  \bibinfo{author}{Linke, N.~M.}, \bibinfo{author}{Sepiol, M.~A.} \&
  \bibinfo{author}{Lucas, D.~M.}
\newblock \bibinfo{title}{High-fidelity quantum logic gates using trapped-ion
  hyperfine qubits}.
\newblock \emph{\bibinfo{journal}{Phys. Rev. Lett.}}
  \textbf{\bibinfo{volume}{117}}, \bibinfo{pages}{060504}
  (\bibinfo{year}{2016}).

\bibitem{Gaebler2016}
\bibinfo{author}{Gaebler, J.~P.} \emph{et~al.}
\newblock \bibinfo{title}{High-fidelity universal gate set for
  ${^{9}\mathrm{Be}}^{+}$ ion qubits}.
\newblock \emph{\bibinfo{journal}{Phys. Rev. Lett.}}
  \textbf{\bibinfo{volume}{117}}, \bibinfo{pages}{060505}
  (\bibinfo{year}{2016}).

\bibitem{McKay2019}
\bibinfo{author}{McKay, D.~C.}, \bibinfo{author}{Sheldon, S.},
  \bibinfo{author}{Smolin, J.~A.}, \bibinfo{author}{Chow, J.~M.} \&
  \bibinfo{author}{Gambetta, J.~M.}
\newblock \bibinfo{title}{Three-qubit randomized benchmarking}.
\newblock \emph{\bibinfo{journal}{Phys. Rev. Lett.}}
  \textbf{\bibinfo{volume}{122}}, \bibinfo{pages}{200502}
  (\bibinfo{year}{2019}).

\bibitem{Meyer2001}
\bibinfo{author}{Meyer, V.} \emph{et~al.}
\newblock \bibinfo{title}{Experimental demonstration of entanglement-enhanced
  rotation angle estimation using trapped ions}.
\newblock \emph{\bibinfo{journal}{Phys. Rev. Lett.}}
  \textbf{\bibinfo{volume}{86}}, \bibinfo{pages}{5870} (\bibinfo{year}{2001}).

\bibitem{Cox2016}
\bibinfo{author}{Cox, K.~C.}, \bibinfo{author}{Greve, G.~P.},
  \bibinfo{author}{Weiner, J.~M.} \& \bibinfo{author}{Thompson, J.~K.}
\newblock \bibinfo{title}{Deterministic squeezed states with collective
  measurements and feedback}.
\newblock \emph{\bibinfo{journal}{Phys. Rev. Lett.}}
  \textbf{\bibinfo{volume}{116}}, \bibinfo{pages}{093602}
  (\bibinfo{year}{2016}).

\bibitem{Hosten2016a}
\bibinfo{author}{Hosten, O.}, \bibinfo{author}{Engelsen, N.~J.},
  \bibinfo{author}{Krishnakumar, R.} \& \bibinfo{author}{Kasevich, M.~A.}
\newblock \bibinfo{title}{Measurement noise 100 times lower than the
  quantum-projection limit using entangled atoms}.
\newblock \emph{\bibinfo{journal}{Nature}} \textbf{\bibinfo{volume}{529}},
  \bibinfo{pages}{505--508} (\bibinfo{year}{2016}).

\bibitem{Bohnet2016}
\bibinfo{author}{Bohnet, J.~G.} \emph{et~al.}
\newblock \bibinfo{title}{Quantum spin dynamics and entanglement generation
  with hundreds of trapped ions}.
\newblock \emph{\bibinfo{journal}{Science}} \textbf{\bibinfo{volume}{352}},
  \bibinfo{pages}{1297--1301} (\bibinfo{year}{2016}).

\bibitem{mottl2012roton}
\bibinfo{author}{Mottl, R.} \emph{et~al.}
\newblock \bibinfo{title}{Roton-type mode softening in a quantum gas with
  cavity-mediated long-range interactions}.
\newblock \emph{\bibinfo{journal}{Science}} \textbf{\bibinfo{volume}{336}},
  \bibinfo{pages}{1570--1573} (\bibinfo{year}{2012}).

\bibitem{leonard2017}
\bibinfo{author}{L{\'e}onard, J.}, \bibinfo{author}{Morales, A.},
  \bibinfo{author}{Zupancic, P.}, \bibinfo{author}{Esslinger, T.} \&
  \bibinfo{author}{Donner, T.}
\newblock \bibinfo{title}{Supersolid formation in a quantum gas breaking a
  continuous translational symmetry}.
\newblock \emph{\bibinfo{journal}{Nature}} \textbf{\bibinfo{volume}{543}},
  \bibinfo{pages}{87--90} (\bibinfo{year}{2017}).

\bibitem{Ozeri2007}
\bibinfo{author}{Ozeri, R.} \emph{et~al.}
\newblock \bibinfo{title}{{Errors in trapped-ion quantum gates due to
  spontaneous photon scattering}}.
\newblock \emph{\bibinfo{journal}{Phys. Rev. A}} \textbf{\bibinfo{volume}{75}},
  \bibinfo{pages}{042329} (\bibinfo{year}{2007}).

\bibitem{Sackett2000}
\bibinfo{author}{Sackett, C.~A.} \emph{et~al.}
\newblock \bibinfo{title}{Experimental entanglement of four particles}.
\newblock \emph{\bibinfo{journal}{Nature}} \textbf{\bibinfo{volume}{404}},
  \bibinfo{pages}{25--259} (\bibinfo{year}{2000}).

\bibitem{Leibfried2003exp}
\bibinfo{author}{Leibfried, D.} \emph{et~al.}
\newblock \bibinfo{title}{Experimental demonstration of a robust, high-fidelity
  geometric two ion-qubit phase gate}.
\newblock \emph{\bibinfo{journal}{Nature}} \textbf{\bibinfo{volume}{422}},
  \bibinfo{pages}{412--415} (\bibinfo{year}{2003}).

\bibitem{Tavis1968}
\bibinfo{author}{Tavis, M.} \& \bibinfo{author}{Cummings, F.~W.}
\newblock \bibinfo{title}{Exact solution for an $n$-molecule---radiation-field
  hamiltonian}.
\newblock \emph{\bibinfo{journal}{Phys. Rev.}} \textbf{\bibinfo{volume}{170}},
  \bibinfo{pages}{379--384} (\bibinfo{year}{1968}).

\bibitem{Molmer1999b}
\bibinfo{author}{M\o{}lmer, K.} \& \bibinfo{author}{S\o{}rensen, A.}
\newblock \bibinfo{title}{Multiparticle entanglement of hot trapped ions}.
\newblock \emph{\bibinfo{journal}{Phys. Rev. Lett.}}
  \textbf{\bibinfo{volume}{82}}, \bibinfo{pages}{1835--1838}
  (\bibinfo{year}{1999}).

\bibitem{Bogoliubov1958}
\bibinfo{author}{Bogoljubov, N.~N.}
\newblock \bibinfo{title}{On a new method in the theory of superconductivity}.
\newblock \emph{\bibinfo{journal}{Il Nuovo Cimento}}
  \textbf{\bibinfo{volume}{7}}, \bibinfo{pages}{794--805}
  (\bibinfo{year}{1958}).

\bibitem{Burd2019}
\bibinfo{author}{Burd, S.~C.} \emph{et~al.}
\newblock \bibinfo{title}{Quantum amplification of mechanical oscillator
  motion}.
\newblock \emph{\bibinfo{journal}{Science}} \textbf{\bibinfo{volume}{364}},
  \bibinfo{pages}{1163--1165} (\bibinfo{year}{2019}).

\bibitem{Seidelin2006}
\bibinfo{author}{Seidelin, S.} \emph{et~al.}
\newblock \bibinfo{title}{Microfabricated surface-electrode ion trap for
  scalable quantum information processing}.
\newblock \emph{\bibinfo{journal}{Phys. Rev. Lett.}}
  \textbf{\bibinfo{volume}{96}}, \bibinfo{pages}{253003}
  (\bibinfo{year}{2006}).

\bibitem{Srinivas2019}
\bibinfo{author}{Srinivas, R.} \emph{et~al.}
\newblock \bibinfo{title}{Trapped-ion spin-motion coupling with microwaves and
  a near-motional oscillating magnetic field gradient}.
\newblock \emph{\bibinfo{journal}{Phys. Rev. Lett.}}
  \textbf{\bibinfo{volume}{122}}, \bibinfo{pages}{163201}
  (\bibinfo{year}{2019}).

\bibitem{Monroe1995}
\bibinfo{author}{Monroe, C.} \emph{et~al.}
\newblock \bibinfo{title}{Resolved-sideband {R}aman cooling of a bound atom to
  the 3{D} zero-point energy}.
\newblock \emph{\bibinfo{journal}{Phys. Rev. Lett.}}
  \textbf{\bibinfo{volume}{75}}, \bibinfo{pages}{4011--4014}
  (\bibinfo{year}{1995}).

\bibitem{Ospelkaus2008}
\bibinfo{author}{Ospelkaus, C.} \emph{et~al.}
\newblock \bibinfo{title}{{Trapped-ion quantum logic gates based on oscillating
  magnetic fields}}.
\newblock \emph{\bibinfo{journal}{Phys. Rev. Lett.}}
  \textbf{\bibinfo{volume}{101}}, \bibinfo{pages}{090502}
  (\bibinfo{year}{2008}).

\bibitem{Ospelkaus2011}
\bibinfo{author}{Ospelkaus, C.} \emph{et~al.}
\newblock \bibinfo{title}{{Microwave quantum logic gates for trapped ions}}.
\newblock \emph{\bibinfo{journal}{Nature}} \textbf{\bibinfo{volume}{476}},
  \bibinfo{pages}{181--184} (\bibinfo{year}{2011}).

\bibitem{Keith2018}
\bibinfo{author}{Keith, A.~C.}, \bibinfo{author}{Baldwin, C.~H.},
  \bibinfo{author}{Glancy, S.} \& \bibinfo{author}{Knill, E.}
\newblock \bibinfo{title}{Joint quantum-state and measurement tomography with
  incomplete measurements}.
\newblock \emph{\bibinfo{journal}{Phys. Rev. A}} \textbf{\bibinfo{volume}{98}},
  \bibinfo{pages}{042318} (\bibinfo{year}{2018}).

\bibitem{Uys2010}
\bibinfo{author}{Uys, H.} \emph{et~al.}
\newblock \bibinfo{title}{Decoherence due to elastic rayleigh scattering}.
\newblock \emph{\bibinfo{journal}{Phys Rev. Lett.}}
  \textbf{\bibinfo{volume}{105}}, \bibinfo{pages}{200401}
  (\bibinfo{year}{2010}).

\bibitem{Babadi2017}
\bibinfo{author}{Babadi, M.}, \bibinfo{author}{Knap, M.},
  \bibinfo{author}{Martin, I.}, \bibinfo{author}{Refael, G.} \&
  \bibinfo{author}{Demler, E.}
\newblock \bibinfo{title}{Theory of parametrically amplified electron-phonon
  superconductivity}.
\newblock \emph{\bibinfo{journal}{Phys. Rev. B}} \textbf{\bibinfo{volume}{96}},
  \bibinfo{pages}{014512} (\bibinfo{year}{2017}).

\bibitem{Qin2019}
\bibinfo{author}{Qin, W.}, \bibinfo{author}{Macr\`{\i}, V.},
  \bibinfo{author}{Miranowicz, A.}, \bibinfo{author}{Savasta, S.} \&
  \bibinfo{author}{Nori, F.}
\newblock \bibinfo{title}{Emission of photon pairs by mechanical stimulation of
  the squeezed vacuum}.
\newblock \emph{\bibinfo{journal}{Phys. Rev. A}}
  \textbf{\bibinfo{volume}{100}}, \bibinfo{pages}{062501}
  (\bibinfo{year}{2019}).

\bibitem{Wineland1998}
\bibinfo{author}{Wineland, D.~J.} \emph{et~al.}
\newblock \bibinfo{title}{{Experimental issues in coherent quantum-state
  manipulation of trapped atomic ions}}.
\newblock \emph{\bibinfo{journal}{J. Res. Natl. Inst. Stand. Technol.}}
  \textbf{\bibinfo{volume}{103}}, \bibinfo{pages}{259--328}
  (\bibinfo{year}{1998}).

\bibitem{Heinzen1990}
\bibinfo{author}{Heinzen, D.~J.} \& \bibinfo{author}{Wineland, D.~J.}
\newblock \bibinfo{title}{{Quantum-limited cooling and detection of
  radio-frequency oscillations by laser-cooled ions}}.
\newblock \emph{\bibinfo{journal}{Phys. Rev. A}} \textbf{\bibinfo{volume}{42}},
  \bibinfo{pages}{2977--2994} (\bibinfo{year}{1990}).

\bibitem{Walls1994}
\bibinfo{author}{Walls, D.~F.} \& \bibinfo{author}{Milburn, G.~J.}
\newblock \emph{\bibinfo{title}{Quantum {O}ptics}}
  (\bibinfo{publisher}{Springer-Verlag}, \bibinfo{address}{Berlin, Germany},
  \bibinfo{year}{1994}).

\bibitem{tellinghuisen2001}
\bibinfo{author}{Tellinghuisen, J.}
\newblock \bibinfo{title}{Statistical error propagation}.
\newblock \emph{\bibinfo{journal}{J. Phys. Chem. A}}
  \textbf{\bibinfo{volume}{105}}, \bibinfo{pages}{3917--3921}
  (\bibinfo{year}{2001}).

\bibitem{Efron1994}
\bibinfo{author}{Efron, B.} \& \bibinfo{author}{Tibshirani, R.~J.}
\newblock \emph{\bibinfo{title}{An introduction to the bootstrap}}
  (\bibinfo{publisher}{CRC press}, \bibinfo{address}{Boca Raton, Florida},
  \bibinfo{year}{1994}).

\bibitem{Turchette2000}
\bibinfo{author}{Turchette, Q.~A.} \emph{et~al.}
\newblock \bibinfo{title}{Decoherence and decay of motional quantum states of a
  trapped atom coupled to engineered reservoirs}.
\newblock \emph{\bibinfo{journal}{Phys. Rev. A}} \textbf{\bibinfo{volume}{62}},
  \bibinfo{pages}{053807} (\bibinfo{year}{2000}).

\bibitem{Brownnutt2015}
\bibinfo{author}{Brownnutt, M.}, \bibinfo{author}{Kumph, M.},
  \bibinfo{author}{Rabl, P.} \& \bibinfo{author}{Blatt, R.}
\newblock \bibinfo{title}{Ion-trap measurements of electric-field noise near
  surfaces}.
\newblock \emph{\bibinfo{journal}{Rev. Mod. Phys.}}
  \textbf{\bibinfo{volume}{87}}, \bibinfo{pages}{1419--1482}
  (\bibinfo{year}{2015}).

\end{thebibliography}
\end{document}